\documentclass[twocolumn,10pt,aps, pra, nofootinbib, superscriptaddress, showkeys,showpacs,longbibliography]{revtex4-1}
\usepackage{graphicx, mathtools, dsfont, mathrsfs, float, braket}
\usepackage[dvipsnames]{xcolor}
\usepackage{epsfig, epstopdf}
\usepackage{ragged2e, booktabs}
\usepackage{amsfonts, amsmath, amsthm} 
\usepackage{wrapfig} 
\usepackage[font=small,labelfont=bf,justification=raggedright, format=plain]{caption}
\usepackage[colorlinks, citecolor = red, urlcolor=blue]{hyperref}

\begin{document}
\title{Universal quantum computing using single-particle discrete-time quantum walk}
\author{Shivani Singh}
\thanks{These authors contributed equally to this work}
\affiliation{The Institute of Mathematical Sciences, C. I. T. Campus, Taramani, Chennai 600113, India}
\affiliation{Homi Bhabha National Institute, Training School Complex, Anushakti Nagar, Mumbai 400094, India}
\author{Prateek Chawla}
\thanks{These authors contributed equally to this work}
\affiliation{The Institute of Mathematical Sciences, C. I. T. Campus, Taramani, Chennai 600113, India}
\affiliation{Homi Bhabha National Institute, Training School Complex, Anushakti Nagar, Mumbai 400094, India}
\author{Anupam Sarkar}
\affiliation{The Institute of Mathematical Sciences, C. I. T. Campus, Taramani, Chennai 600113, India}
\affiliation{Homi Bhabha National Institute, Training School Complex, Anushakti Nagar, Mumbai 400094, India}
\author{C. M. Chandrashekar}
\email{chandru@imsc.res.in}
\affiliation{The Institute of Mathematical Sciences, C. I. T. Campus, Taramani, Chennai 600113, India}
\affiliation{Homi Bhabha National Institute, Training School Complex, Anushakti Nagar, Mumbai 400094, India}

\begin{abstract}
Quantum walk has been regarded as a primitive to universal quantum computation.  By using the operations required to describe the single particle discrete-time quantum walk on a position space we demonstrate the realization of the universal set of quantum gates on two- and three-qubit systems. The idea is to utilize the effective Hilbert space of the single qubit and the position space on which it evolves 
in order to realize multi-qubit states and universal set of quantum gates on them. Realization of many non-trivial gates and engineering arbitrary states is simpler in the proposed quantum walk model when compared to the circuit based model of computation. We  will also discuss the scalability of the model and some propositions for using lesser number of qubits in realizing larger qubit systems.
\end{abstract}

\maketitle
\section{\label{sec:intro}Introduction}
Quantum walk\,\cite{GVR58,RPF86,ADZ93,DAM96, FG98}, a quantum mechanical analogue of classical random walk has been a basis for many quantum algorithms\,\cite{YKE08,JK03,ESV12,ANAV17,IKS05,CMC20} and schemes for quantum simulations\,\cite{MRL08, SF92,KTB10, C11, C13, MMC17,CAC20}.  The dynamics of quantum walk have been described in several ways,  however, they can be broadly classified under the two of the most distinct and prominent categories, the continuous-time and discrete-time quantum walks.  Engineering quantum gates and realizing the set of universal quantum gates has been shown using both these forms of quantum walks\,\cite{AMC09,LCE10,HSK16}. This means that any problem that can be solved on a quantum computer can also be solved using quantum walks.  The one-dimensional discrete-time quantum walk has also been used to engineer arbitrary qudit states \cite{IMG17}. It has been experimentally implemented on a linear optical system which uses the orbital angular momentum degree of freedom of single photon states to represent the particle \cite{GPE19}. The idea is to increase the control over the dynamics of walk by using appropriate evolution operators and thus driving the particles' state towards the desired qudit state. Theoretically, this technique can be used to prepare any high-dimensional quantum state and experimentally, a six-dimensional qudit state has been prepared and measured. All this highlights the versatility of quantum walks.  
The scheme for quantum computation presented here is based on controlled dynamics of the walk with the help of appropriate position dependent evolution operators and has a scope of designing an architecture for quantum processor using quantum walks.\\
\noindent
Experimental demonstration of an eighteen-qubit entangled state ($2^{18}$ possible states) from six individual photons by simultaneously using three degrees of freedom\,\cite{WLH18} and demonstration of flexible two-qubit quantum computation from a single photon\,\cite{YSZ20}, highlight the potential power of associated Hilbert space with the photon in realizing higher number of qubits.  This, along with the ability to engineer the quantum walk dynamics serve as a strong motivation for us to explore a resourceful way to use lesser number of particles to realize an entangled state of a bigger system. Thus, in this work, we focus on exploring the power of single particle discrete-time quantum walk in order to realize a multi-qubit computational model. \\
\noindent
One of the main criteria for a system to be considered as a suitable candidate for  universal quantum computation is its ability to realize a universal set of quantum gates.  A set of gates is called universal for quantum computation if it can reproduce an approximation of any $n \geq 1$-qubit unitary operator to an arbitrary accuracy on a quantum circuit. In general, the universal set of gates are $\{P, H, CNOT \}$\,\cite{NC00}, where phase ($P$) and Hadamard ($H$) are single qubit gates and controlled-NOT ($CNOT$) is a two qubit gate.\\ 
\noindent
Quantum computing has been shown on both forms of quantum walks, i.e., continuous-time\,\cite{AMC09} and discrete-time quantum walks\,\cite{LCE10}, where the position space of the particle represents quantum wires. They give a way of programming a quantum computer rather than modelling or mimicking one and hence do not exhibit a potential towards designing a physical architecture. On the other hand, our model, based on the discrete-time quantum walk, gives a physical and logical building block to model a quantum computer on lattice based system or photonic systems. The scheme presented here maps the position basis state to the qubit state and then performs quantum computation by mimicking gates using evolution unitaries.\\
\noindent
In this work we consider a single physical particle (qubit) with positional degrees of freedom to mimic the computational basis of a multi-qubit system. By using a set of operations used to describe the dynamics of quantum walk we show the realization of universal set of gates and a controlled-Z gate on two- and three-qubit systems using the single particle quantum walk in a position space consisting of two and four points, respectively (Sec.\,\ref{sec:univcomp}). In our scheme, the ability of the particle to hop between different points in the position space in superposition makes the realization of many non-trivial gates much simpler when compared to the circuit model of computation.  We demonstrate this by presenting the scheme for realization of a simple three qubit circuit and creation of a GHZ-state in Sec.\,\ref{circuit}. Scalability of the model, its practical relevance and some propositions for using lesser number of qubits in realizing larger qubit system are presented in Sec.\,\ref{sec:scaling}. 
We conclude with our remarks in Sec.\,\ref{sec:conc}. \\  

\noindent
\section{\label{sec:univcomp} Quantum computation via discrete-time quantum walk}

\subsection{\label{sec:dtqw} Discrete-time quantum walk}
\noindent
The dynamics of the one dimensional discrete-time quantum walk on a line are described by a particle with two internal degrees of freedom, which is defined on a combined Hilbert space $\mathcal{H}_w=\mathcal{H}_c \otimes  \mathcal{H}_p$. The coin Hilbert space, $\mathcal{H}_c  = span\{\ket{0},  \ket{1}\}$ represents the internal coin states and  position Hilbert space, $ \mathcal{H}_p  = span\{\ket{l}\}$, $l  \in \mathbb{Z}$ represents the number of position states available to the particle.\\
\noindent
Evolution of each step in the walk is defined by the action of the unitary quantum coin operation followed by the position shift operation. The general form of the quantum coin operator is a non-orthogonal unitary\,\cite{CSL08} which acts only on the coin space, and is given by, 
\begin{equation}
\hat{C}(\tau,\xi,\zeta,\theta) = e^{i\tau}
\begin{bmatrix}
~~e^{i \xi}\cos(\theta) & ~~~~ e^{i \zeta}\sin(\theta) \\
e^{-i \zeta}   \sin(\theta) & - e^{-i \xi} \cos(\theta) 
\end{bmatrix}.
\label{qcoin}
\end{equation}
The position shift operators, $\hat{S}_-$ and $\hat{S}_+$ translate the particle to the left and right, respectively, conditioned on the internal state of the particle. They are of the form,

\begin{align}
\hat{S}_{-}^{k}  &=  \sum_{\substack{l\in\mathbb{Z} \\ j }}  \Big [\ket{k}\bra{k}
\otimes   \ket{l-1}\bra{l}+\ket{j \neq k}\bra{j \neq k} \otimes
\ket{l}\bra{l} \Big ]   \nonumber \\
\hat{S}_{+}^{j}  &= \sum_{\substack{l\in\mathbb{Z} \\ k }}  \Big [ \ket{k \neq j}\bra{k \neq j}
\otimes  \ket{l}\bra{l}+\ket{j}\bra{j} \otimes \ket{l+1}\bra{l} \Big ].
\label{qshift}
\end{align}


Here, $\ket{k}$ and $\ket{j}$ are the basis states of coin Hilbert space $\mathcal{H}_c$, i.e., $\ket{k},\ket{j} \in \{\ket{0}, \ket{1} \}$.  The operator $W_{ss} = (\hat{S}_{+}^{1} \hat{C}(\tau_2,\xi_2,\zeta_2,\theta_2) \otimes \mathbb{I}_p)(\hat{S}_{-}^{0} \hat{C}(\tau_1,\xi_1,\zeta_1,\theta_1) \otimes \mathbb{I}_p$) implements one step of split-step quantum walk\,\cite{KTB10, MC16} and the operator $W_{d} = (\hat{S}_{\pm}^{1}\big(\hat{C}(\tau,\xi,\zeta,\theta) \otimes \mathbb{I}_p\big)$ implements one step of directed quantum walk (conditioned on the state $\ket{1}$)\,\cite{AM05, CB14, SD09}, a variant of discrete-time quantum walk which results in non-zero probability at all of the position space it spans through while walking. The set of operators $\Big\{ \hat{S}_{\pm}^{0}, \hat{S}_{\pm}^{1}, \hat{C}(\tau,\xi,\zeta,\theta)\Big\}$ along with the identity operator $\hat{S} = \mathbb{I}$ can be considered a generic set of operators that describes the quantum walk. We will use this set of operators for the realization of universal quantum gates on a two- and three-qubit system by mapping the position space to the computational basis.  \\
\noindent
\subsection{\label{sec:univgates} Universal quantum gates}
\noindent
The universal set of quantum gates for quantum computation comprises of two single qubit gates-- Phase gate ($P$) and Hadamard gate ($H$) and one two-qubit gate-- controlled-NOT gate ($CNOT$), i.e.,
\begin{equation}
\begin{split}
U = & \bigg \{ P, H, CNOT \bigg \} \\
= & \Bigg \{ \begin{bmatrix}
1 & 0 \\
0 & e^{i\phi}
\end{bmatrix},  \frac{1}{\sqrt{2}} \begin{bmatrix}
1 & ~~1 \\
1 & -1
\end{bmatrix},     \begin{bmatrix}
1 & 0 & 0 & 0\\
0 & 1 & 0 & 0\\
0 & 0 & 0 & 1\\
0 & 0 & 1 & 0
\end{bmatrix} \Bigg \}.
\end{split}
\end{equation}
The action of phase gate is given by, 
$P \ket{0} = \ket{0}$ and $ P \ket{1} = e^{i\phi} \ket{1}$. The action of Hadamard gate is given by $H \ket{0} = \frac{1}{\sqrt{2}} (\ket{0} + \ket{1})$ and $H \ket{1} = \frac{1}{\sqrt{2}} (\ket{0} - \ket{1})$. Similarly, the action of CNOT gate is given by,
$CNOT \ket{00} = \ket{00}$, $CNOT \ket{01} = \ket{01}$, $CNOT \ket{10} = \ket{11}$, and $CNOT \ket{11} = \ket{10}$, where the first qubit is the control bit and the second qubit is the target bit.\\
\noindent
\section{\label{sec:Method} Methods}
\noindent
In the quantum walk scheme, a gate operation is performed with the help of the evolution operations and initial state can be defined by the initial state of the particle. The direction of the quantum walk during the circuit operation is defined by the directed walk evolution operator. The particle remains at the initial position state with certain probability based on the form of coin operator, and moves with certain probability in either forward or backward direction based on the shift operation given by equation\,\eqref{qshift}.
\noindent
\section{Results}
\subsection{\label{sec:setup} Quantum walk set-up for computation on two and three qubit system}
\begin{figure}
	\centering
	\includegraphics[width=0.8\linewidth]{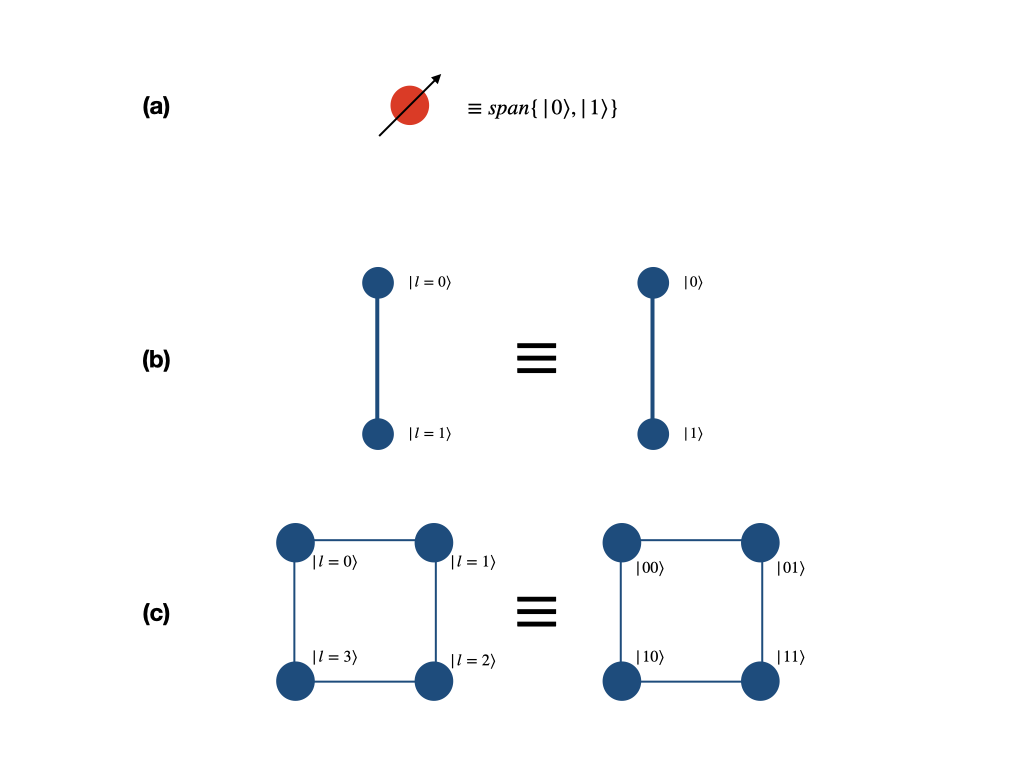}
	\caption{(a) Shows an illustration of a real qubit state. (b)  Shows a mapping between the two states of position Hilbert space in one-dimension  to the computational basis of second qubit in two qubit system, and (c) shows a mapping between the position Hilbert space in one-dimensional closed quantum walk state to the computational basis of the second and third qubits in the three qubit system. These two graphs for quantum walk form the building blocks for the scheme to perform computation using quantum walks.} 
	\label{Mapping}
\end{figure}

\noindent
The quantum walk based quantum computation scheme proposed herein uses a directed shift operation with a position dependent coin operator to realize the gate operation. To perform the operations of universal set of gates on two qubit system, the particle will execute a quantum walk on an open graph of two vertices such that particle itself will act as first qubit with two internal degrees of freedom, $span \{\ket{0},  \ket{1}\}$ representing the state of the first qubit. The second qubit will be represented by the position space on which the walk is performed as shown in the Fig.\,\ref{Mapping}-(b). Similarly, for three qubit case, first qubit is represented by the particle's internal degree of freedom and the remaining two qubits states are mapped on the position space. The position space is a two dimensional closed graph with four vertices and four edges, $span \{\ket{00}, \ket{01}, \ket{11}, \ket{10} \}$ as shown in the Fig.\,\ref{Mapping}-(c), on which gate operations are performed. \\

\noindent
In the previously known schemes\,\cite{AMC09, LCE10},  position space (computational basis state) was used as a quantum `wire' and gates required for universality were then attached to these wires. The flow of the computation from input to output was represented as a quantum walk on these wires. The computational basis states thus represented wires rather than qubits and thus these models did not admit a physical architecture straightaway. In the scheme presented here, however, the computation basis represents the qubit and the universal gates are mimicked with the help of controlled evolution operators. As a consequence, the proposed scheme is closer to physical architecture.  Direction of the flow (role of wire), is given by the shift operators (evolution operator) of directed quantum walk type. This scheme has a scope of being used for quantum computation on a system with access to position basis states e.g., photonic or lattice based system. Quantum walks in position space with sufficient control over dynamics have already been experimentally implemented for different purposes\,\cite{GPE19, WLH18} and it favours our scheme to be used in future for quantum computation due to the fact that it is simpler and straightforward.\\

\subsection{\label{sec:gates} Quantum gates on discrete-time quantum walk}
\noindent
Below we describe the mapping of the single particle quantum walk system to the computational basis of the two-qubit and three-qubit system. The arrow shows the forward (positive) direction of the particle. We further present the appropriate combination of shift and coin operations that describes the quantum walk and effectively implements the universal set of gates on the computational basis. In order to realize the two-qubit gates using single particle quantum walk, the mapping of the physical system to the gate implementation is done by using the coin space as the first qubit and the two points in the position as the second qubit. Similarly, mapping for the three-qubit system is done using a single particle on a four points in position space. Since we shall use only a single qubit in our scheme, most of the operations described in the rest of this work are essentially position-specific operations in the particle Hilbert space.  \\
\subsubsection{\label{sec:phase} Phase Gate} 
\noindent
To implement a phase gate on a quantum walk system we need only a position dependent coin operation and thus the shift operator takes the form of identity operator. For both the two-qubit and three-qubit systems, when the computational basis of the particle in position space is in the desired two- or three qubit-state, phase gate on the first (real) qubit (particle) can be applied by using phase operator as quantum coin operation, and identity operator in the position space,
\begin{equation}
\label{eqPhaseOp}
P = \begin{bmatrix}
1 & 0 \\ 
0 & ~~e^{i\phi}
\end{bmatrix}\otimes \mathbb{I}_P.
\end{equation}
\begin{figure}
	\centering
	\includegraphics[width=0.4\linewidth]{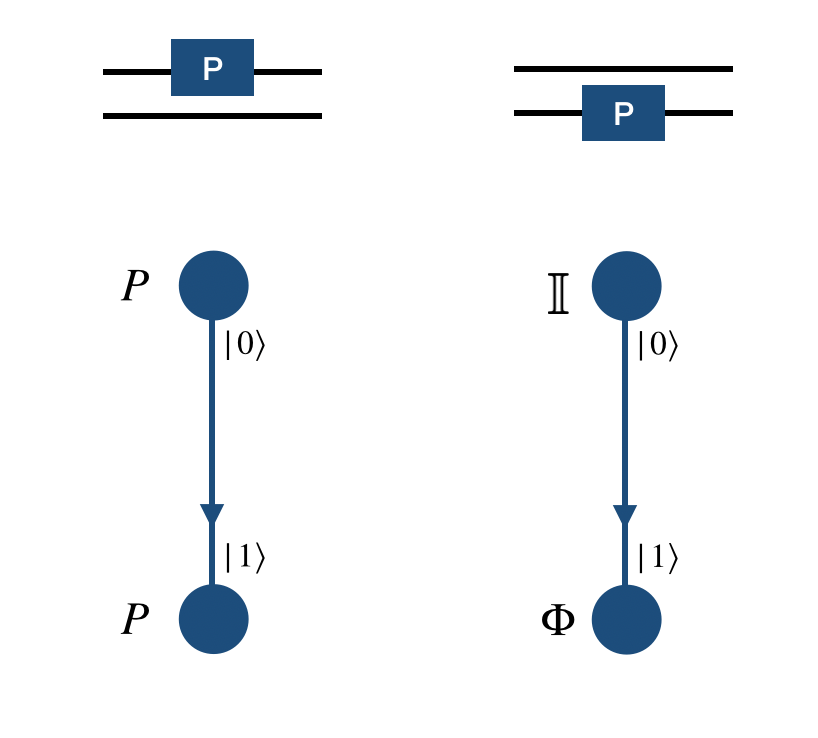}%
	\caption{Schematic illustration for the realization of phase gate on the computational basis of two-qubit system using a single particle quantum walk on two point position space using position dependent coin operation on the real qubit.}
	\label{phase2}
\end{figure}
\begin{figure}
	\centering
	\includegraphics[width=0.8\linewidth]{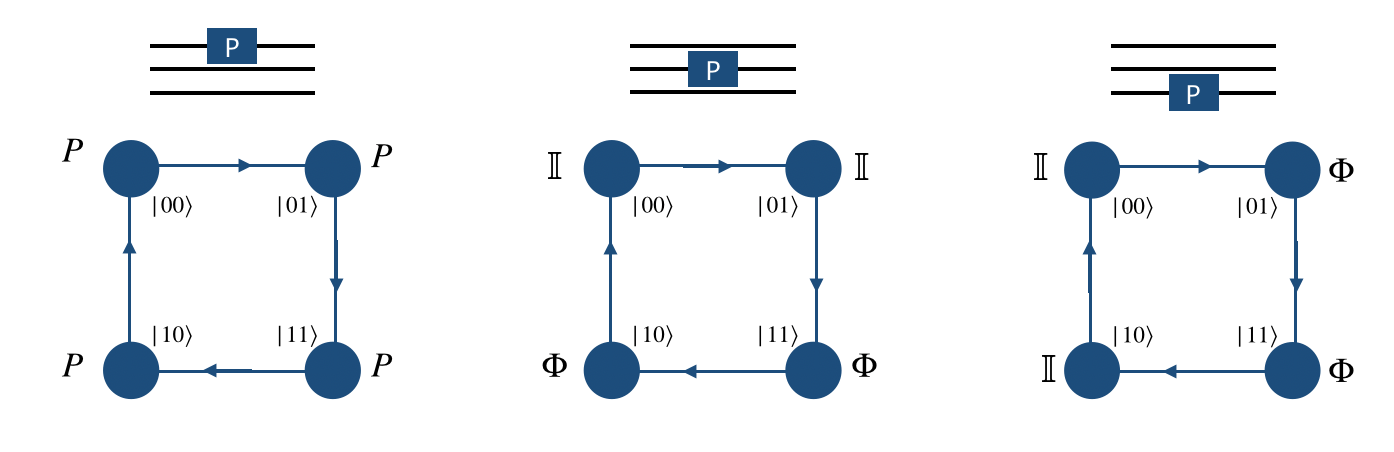}
	\caption{Schematic illustration for the realization of the phase gate on the computational basis of three-qubit system using a single particle quantum walk on four point position space using the position dependent coin operation on the real qubit.}
	\label{phase3}
\end{figure}

\noindent
Applying a phase operator to the second and third qubits requires the implementation of two different types of position specific identity operator separated by a phase,
\begin{equation}
\label{eqPhaseIden}
\Phi = e^{i\phi}\begin{bmatrix}
1 & 0 \\ 0 & 1
\end{bmatrix}\otimes \mathbb{I}_p  ~~~ \mbox{and} ~~~  \mathbb{I} = \mathbb{I} \otimes \mathbb{I}_p .
\end{equation}
In Fig.\,\ref{phase2} we illustrate the mapping of two states in position space to the computation basis of the second qubit in the two-qubit system. The position dependent coin operation on the particle, i.e., $\mathbb{I}$ and $\Phi$ on the space labeled $|0\rangle$ and $|1\rangle$, respectively, will implement the phase gate on the second qubit. Similarly, in Fig.\,\ref{phase3}, we illustrate the mapping of four points in position space to the computation basis of the second and third qubits in the three-qubit system. The position dependent coin operations $\mathbb{I}$ and $\Phi$ on the relevant state in the position space, as shown in Fig.\,\ref{phase3} will implement the phase gate on the second and third qubits.\\ 
\subsubsection{\label{sec:hadamard} Hadamard Gate}
\noindent
Hadamard operation on the first qubit, i.e., the coin state of the particle is given by the evolution operation, 
\begin{align}
\label{eqW}
W = \mathbb{I}(\hat{C}(0,0,0,\pi/4) \otimes \mathbb{I}_p) \equiv \hat{H}_1,
\end{align}
which is a coin operation $\hat{C}(0,0,0,\pi/4)$ as in equation\,\eqref{qcoin} on particle state followed by identity operation on position space as shift operation. The subscript of $\hat{H}$ represents the qubit on which the Hadamard operation is performed.
\begin{figure}
	\centering
	\includegraphics[width=0.4\linewidth]{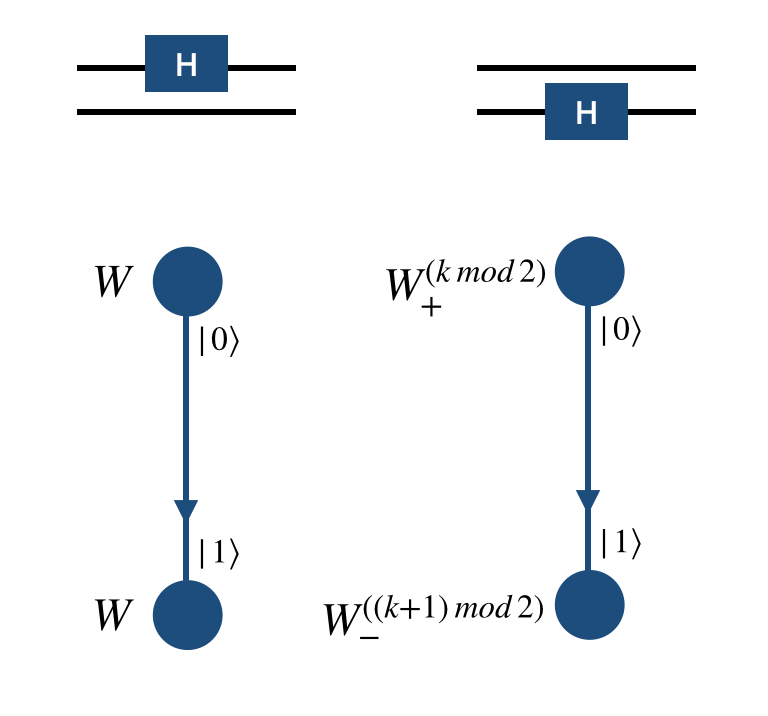}
	\caption{Schematic illustration of Hadamard operation on the computational basis of a two-qubit system using the $W$ operator (equation\,(\ref{eqW})) on a single particle quantum walk on two-point position space. On the first qubit it is the Hadamard operation and on the second qubit Hadamard operation is realized using the $\sigma_x$ and $\mathbb{I}$ operators.}
	\label{Hada1}
\end{figure}
\begin{figure}
	\centering
	\includegraphics[width=0.9\linewidth]{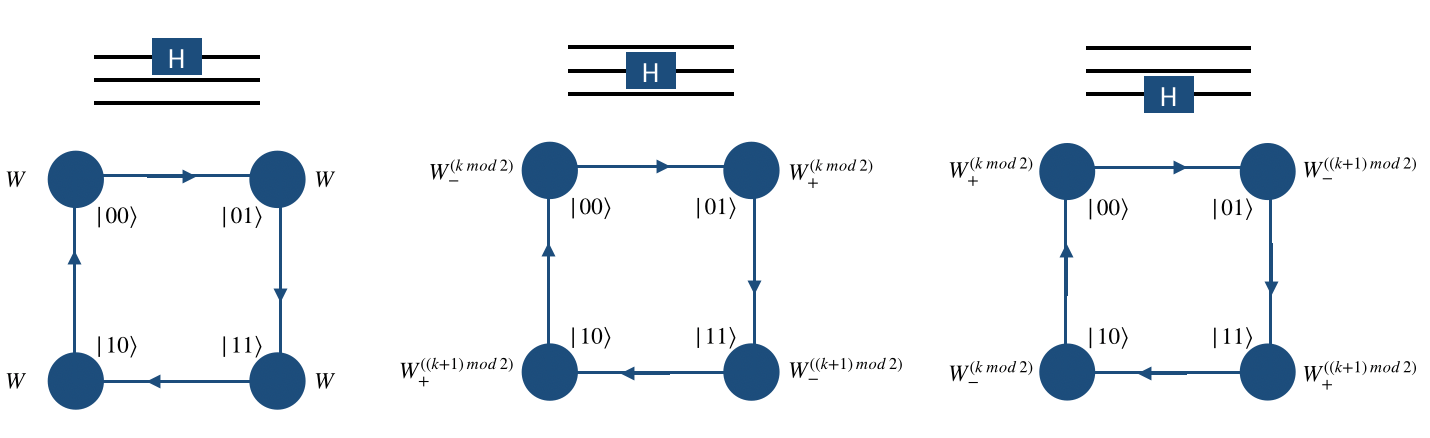}
	\caption{Schematic illustration of the Hadamard operation on the computational basis of the three-qubit system using position dependent quantum walk operators. The form $W$ of the operators involved in realization of Hadamard operation on second and third qubit are in equation\,(\ref{eqWplusMinus}).}
	\label{Hada2}
\end{figure}

\noindent
Hadamard operation on second and third qubits in this computational basis using the single particle quantum walk can be performed by evolving the coin state of the particle in superposition of position space using the different combination of shift operations $\{S_{-}^{k},S_{+}^{j}\} $ as given in equation\,\eqref{qshift}, where $\{\ket{k},\ket{j}\}$ are the coin states, Pauli operations $\sigma_x$ and $\sigma_z$, and Hadamard operation $\hat{H}$ on the coin space of the particle. The quantum walk operations to realize Hadamard operation on second and third qubit in computational basis take the form, \\
\begin{align}
\label{eqWplusMinus}
W_{+}^{0} \ket{k}\otimes \ket{m} &= \Big[\sigma_x^{m}S_{+}^{k}(\sigma_x \otimes \mathbb{I})\Big] \nonumber \\
W_{+}^{1} \ket{k}\otimes \ket{m} &= \Big[\sigma_x^{m}S_{+}^{k}(\sigma_z \otimes \mathbb{I})\Big] \nonumber \\
W_{-}^{0} \ket{j}\otimes \ket{n} &= \Big[\sigma_x^{n}S_{-}^{j} (\sigma_x \otimes \mathbb{I})\Big]  \nonumber \\ 
W_{-}^{1} \ket{j}\otimes \ket{n} &= \Big[\sigma_x^{n}S_{-}^{j} (\sigma_z \otimes \mathbb{I})\Big] ,
\end{align}
%
where, $\sigma_x^{m} = \sigma_x \otimes \ket{m}\bra{m} + \mathbb{I}_c \otimes \sum_{l \neq m} \ket{l}\bra{l}$ and $\ket{m}$ is the initial position state of the particle. In the equation\,\eqref{eqWplusMinus}, Hadamard operator is the coin operation on the particle's coin space for all initial states followed by the conditional shift operator on position space.\\
\noindent
Figs.\,\ref{Hada1} and \,\ref{Hada2} shows the mapping of the states of position space to the computational basis on the second qubit of two-qubit system, and second and third qubits of the three-qubit system, respectively. From the mapping shown in Fig.\,\ref{Hada1}, it is possible to realize Hadamard operation on second qubit in two-qubit system, using $H_2\ket{k0} \equiv W_{+}^{(k \mod 2)}(\hat{H} \otimes \mathbb{I}) \ket{k,l=0}$ and $H_2\ket{k1} \equiv W_{-}^{((k+1) \mod 2)}(\hat{H} \otimes \mathbb{I}) \ket{k,l=1}$ where $\ket{l}$ is the position basis state and $\ket{k}$ is the coin basis state.\\
\noindent
Similarly, as shown in the Fig.\,\ref{Hada2}, one can realize the Hadamard operation on the second and third qubit of the three-qubit system using operations,
\begin{align} \label{HadaEq2}
H_2 \ket{k00} &\rightarrow W_{-}^{(k \mod 2)}(\hat{H} \otimes \mathbb{I}) \ket{k,l=0}, \nonumber \\
H_2 \ket{k01} &\rightarrow W_{+}^{(k \mod 2)}(\hat{H} \otimes \mathbb{I}) \ket{k,l=1}, \nonumber \\
H_2 \ket{k11} &\rightarrow W_{-}^{((k+1) \mod 2)}(\hat{H} \otimes \mathbb{I}) \ket{k,l=2}, \nonumber \\
H_2 \ket{k10} &\rightarrow W_{+}^{((k+1)\mod 2)}(\hat{H} \otimes \mathbb{I}) \ket{k,l=3}
\end{align}
and
\begin{align} \label{HadaEq3}
H_3 \ket{k00} &\rightarrow W_{+}^{(k\mod 2)}(\hat{H} \otimes \mathbb{I}) \ket{k,l=0}, \nonumber \\
H_3 \ket{k01} &\rightarrow W_{-}^{((k+1)\mod 2)}(\hat{H} \otimes \mathbb{I}) \ket{k,l=1}, \nonumber \\
H_3 \ket{k11} &\rightarrow W_{+}^{((k+1)\mod 2)}(\hat{H} \otimes \mathbb{I}) \ket{k,l=2}, \nonumber \\
H_3 \ket{k10} &\rightarrow W_{-}^{(k\mod 2)}(\hat{H} \otimes \mathbb{I}) \ket{k,l=3}
\end{align}
where $\ket{k}$ is the coin basis state of the particle, given as $span\{\ket{0},\ket{1}\}$. Here $l$ are the labels to the points in the position space in a clockwise direction, as illustrated in Fig.\,\ref{Mapping}-(c).\\

\subsubsection{\label{sec:cnot} Controlled-NOT Gate}
\begin{figure}
	\centering
	\includegraphics[width=0.4\linewidth]{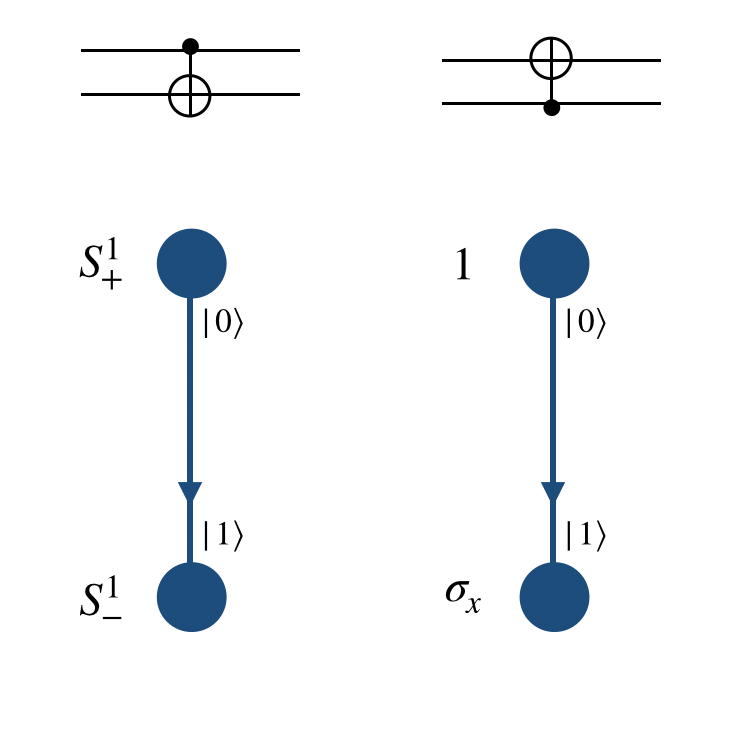}
	\caption{Schematic illustration of the controlled-NOT gate on the computational basis of two-qubit system using a single particle quantum walk on two point position space using position-dependent coin operation on the real qubit. Form of the shift operator $S^1_{\pm}$ is given in equation\,\eqref{qshift}}
	\label{CNOT2}
\end{figure}

\begin{figure}
	\centering
	\includegraphics[width=0.8\linewidth]{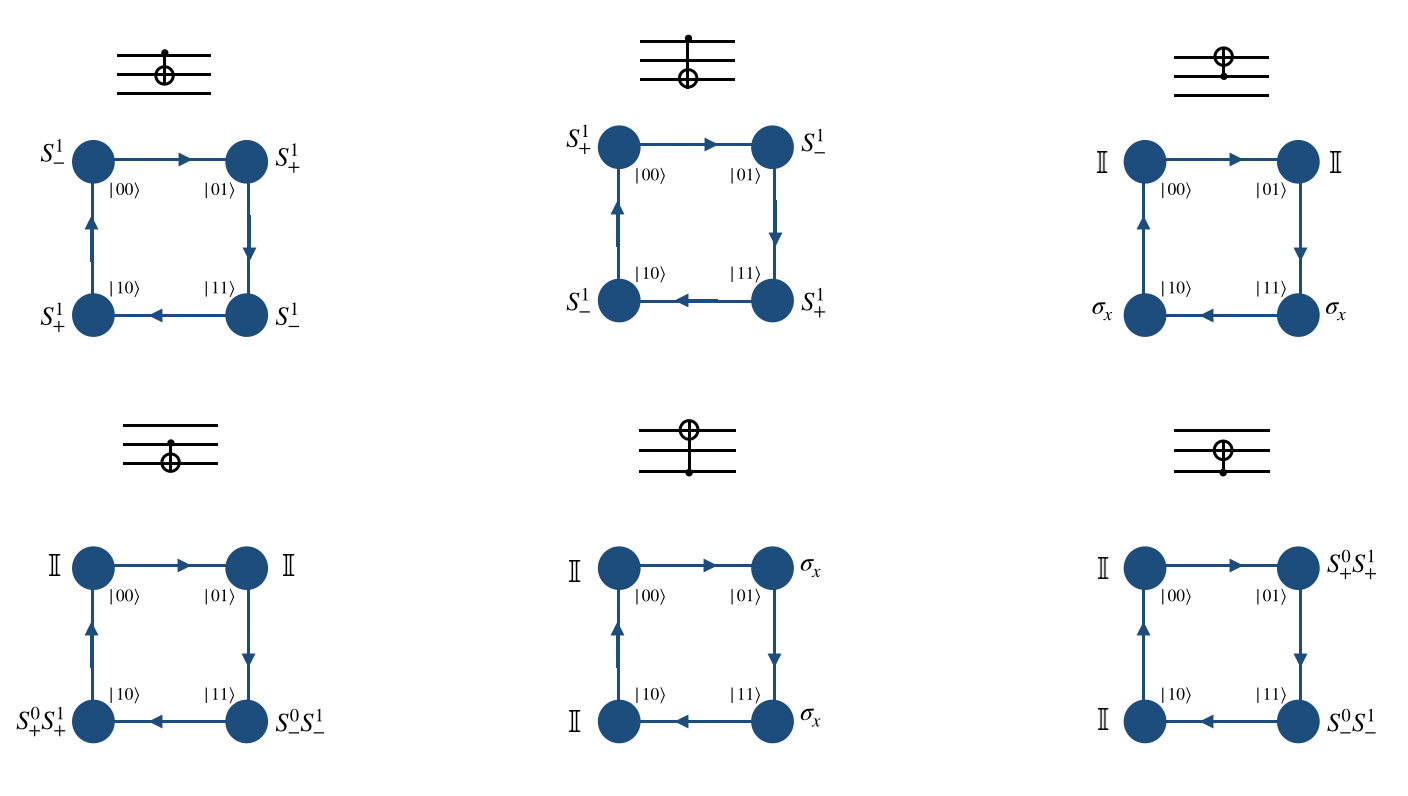}
	\caption{Schematic illustration of the controlled-NOT gate on the computational basis of three-qubit system using position-dependent quantum walk operators. Form of the shift operator $S^j_{\pm}$ is given in equation\,\eqref{qshift}.}
	\label{CNOT3}
\end{figure}

\noindent
This gate can be engineered by evolving the state of the particle using evolution operator which consists of identity as coin operator followed by a position dependent shift operators. The shift operator can be either $S_{+}^{1}$ or $S_{-}^{1}$ as given in equation\,\eqref{qshift} when the coin state is the control and position state (corresponding computational basis) is the target. But when the particle, i.e., coin state is the target, and position state is the control qubit, then the position-dependent coin operation $\hat{C}(0,0,0,\pi/2) \equiv \sigma_x$ followed by the identity shift operator will give controlled-NOT operation implementation on computational basis using single particle. This is schematically illustrated for the two-qubit system in Fig.\,\ref{CNOT2}. \\
\noindent
When second qubit is the control and third qubit is the target in the computational basis of the three-qubit system, the position dependent conditional shift operator $S_{+}^{0}S_{+}^1$ and $S_{-}^{0}S_{-}^{1}$ on the position space with identity operator on coin space implements the CNOT-gate. A similar architecture may be designed for third qubit as control and second as target. These implementations and the corresponding single particle quantum walk operators are schematically illustrated in Fig.\,\ref{CNOT3}. 

\subsubsection{\label{sec:toffoli} Toffoli Gate}
\begin{figure}
	\centering
	\includegraphics[width=0.8\linewidth]{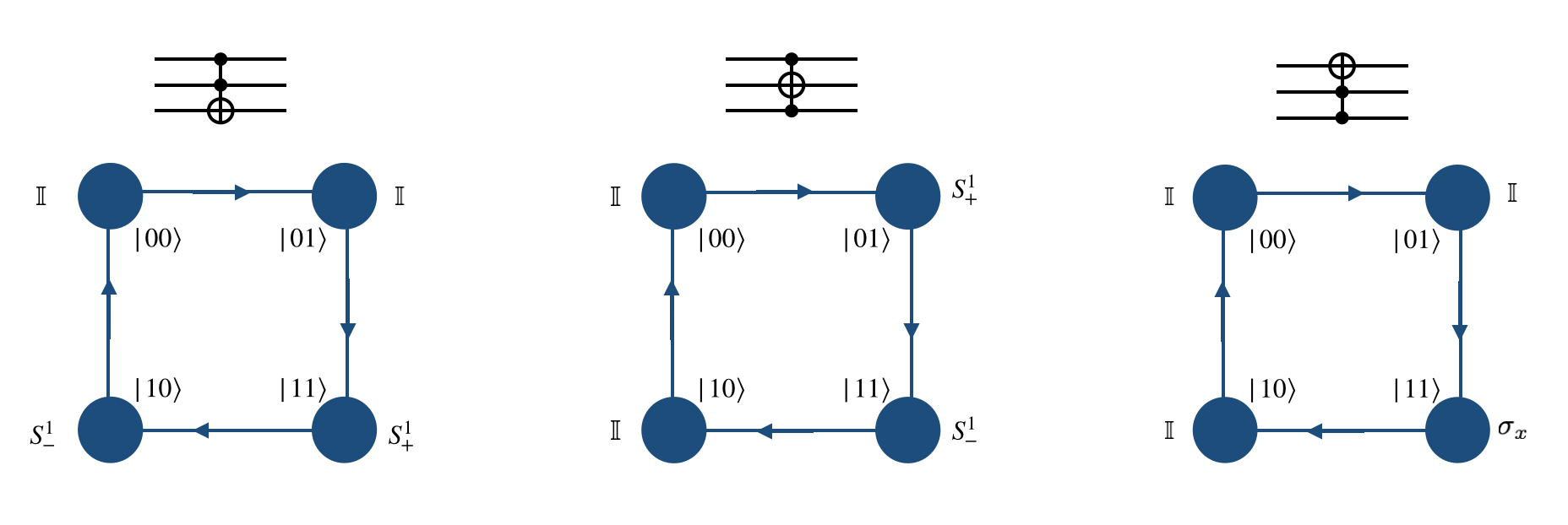}
	\caption{Schematic illustration of the Toffoli gate on a three-qubit system using position dependent quantum walk operators.}
	\label{Toffoli}
\end{figure}

\noindent
This gate can only be realized for a system with three or more qubits. We demonstrate a possible realization of this gate for a three-qubit system. When the first and second qubits are the controls and the third is the target, realization of this gate simply requires conditional shift operations, given by the shift operator $S^1_\pm$, as defined in equation\,\eqref{qshift}. The shift operators are to be applied on certain position basis states only, and other position basis states are simply operated upon by the identity operator. The corresponding scheme is schematically illustrated in Fig.\,\ref{Toffoli}.

\noindent
The Fredkin gate is a controlled swap operation, and closely resembles the Toffoli gate in its implementation. In case when the first or third qubit are the target and other one is the control qubit, this operations for its realization can be worked out exactly the same way as described for the corresponding Toffoli gate.

\subsubsection{\label{sec:czgate}  Controlled-Z gate}
\begin{figure}[!h]
	\centering
	\includegraphics[width=0.85\linewidth]{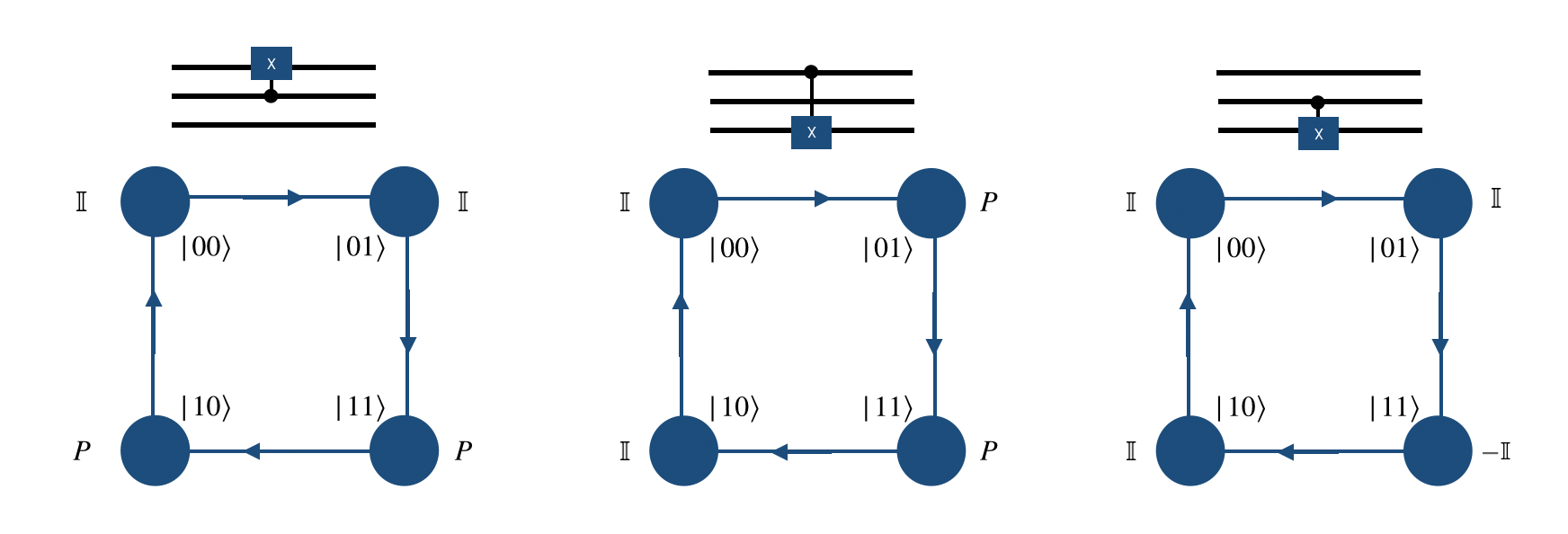}
	\caption{Schematic illustration of the implementation of the controlled-Z gate on a three-qubit system. The operation $P$ is a phase operation of the form $(e^{i\pi}\mathbb{I}\otimes\mathbb{I}_p) $. }
	\label{CZ3}
\end{figure}

\noindent
In a two-qubit system, controlled-Z  gate is closely related to the phase gate, and is implemented exactly like the phase gate applied to the second qubit in the position space, as illustrated in Fig.\,\ref{phase2}. The only distinction in the realization of these two gates is that in the controlled-Z gate, the parameter $\phi$ is fixed, so that $\phi=\pi$.\\
\noindent
In a three-qubit system, this gate can be implemented by using the position dependent application of the phase operator $P$ on some position basis states and identity operator on the others. It is also observed that the implementation of this gate is symmetric, i.e., the implementation of the gate between the $i^{th}$ and $j^{th}$ qubits is the same as the implementation between the $j^{th}$ and $i^{th}$ qubits, where $i,j = {1,2,3}$, and $ i \neq j$. As in the case of a two-qubit system, the parameter $\phi$ is fixed to $\pi$.\\
\noindent
In case the gate is applied between the second and third qubits, the scheme can be implemented by using just two kinds of identity operators, separated by a phase of $\pi$. The identity with the phase $e^{i\pi}\mathbb{I}\otimes\mathbb{I}_p$ is applied only to one position state, whereas all the other states are acted upon by the identity operator $\mathbb{I}\otimes \mathbb{I}_p$. This is schematically illustrated in Fig.\,\ref{CZ3}.

\section{\label{sec:Implementation} Implementing simple circuits and scalability}
\subsection{\label{circuit} Circuit implementation on quantum walk based computation setup}
\begin{figure}
	\centering
	\includegraphics[width=0.7\linewidth]{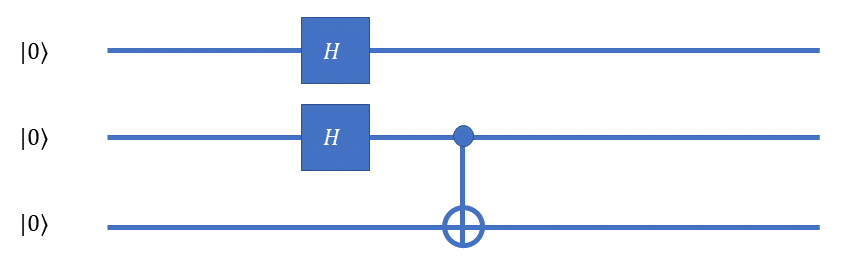}
	\includegraphics[width=0.7\linewidth]{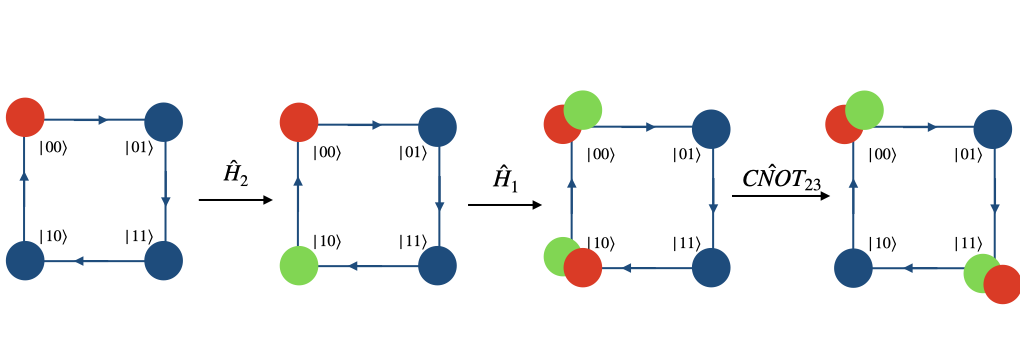}
	\caption{Quantum circuit on three qubit system and equivalent quantum walk scheme to implement same circuit is illustrated. Red circle represents $\ket{0}$ of the real particle and green circle represents $\ket{1}$ of the real particle. The input state is $\ket{\Psi}_{in} = \ket{000}$ and output state is a superposition of four states $\ket{\Psi}_{out} = \frac{1}{2}\Big( \ket{000} + \ket{011} + \ket{100} + \ket{111} \Big)$. $CNOT_{23}$ is a position dependent shift operation given by $\hat{S}^{0}_{-}\hat{S}^{1}_{-}$ at position $\ket{01}$ and identity at other states.}
	\label{cktmodel}
\end{figure}

\noindent
Any two or three qubit circuit can be implemented very easily on this scheme. In Fig.\,\ref{cktmodel}, a simple three qubit circuit and the quantum walk scheme to implement those same gates to get the same output result is shown. The input state for this circuit is $\ket{\Psi}_{in} = \ket{000}$ and the output is $\ket{\Psi}_{out} = \frac{1}{2}\Big( \ket{000} + \ket{011} + \ket{100} + \ket{111} \Big)$. The quantum walk-based scheme can implement the circuit shown in three steps. The first step would be the coin operation followed by shift operation of the form $W^0_-$ to get Hadamard on the second qubit, the second step would be coin operation $\hat{C(0,0,0,\pi/4)}$ followed by identity shift operation to get Hadamard on first qubit and the third step would be identity on coin state followed by position dependent shift operation $S^0_- S^1_+$ on position state $\ket{10}$ to get CNOT-operation, where the coin operator $\hat{C}$ has been defined in equation\,\eqref{qcoin}, the $S^1_+$ in equation\,\eqref{qshift} and $W^{0}_{-}$ in equation\,\eqref{HadaEq2}.
\begin{figure}
	\centering
	\includegraphics[width=0.7\linewidth]{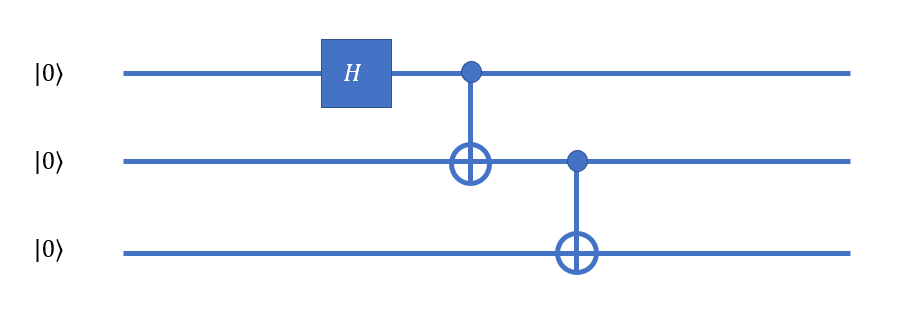}
	\includegraphics[width=0.7\linewidth]{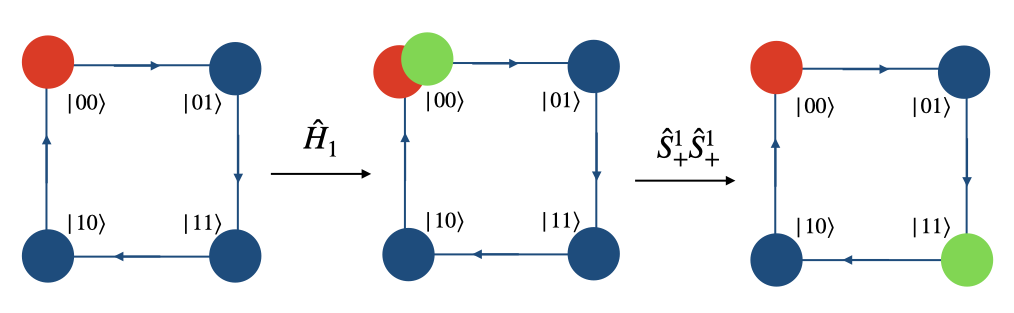}
	\caption{Quantum circuit to create GHZ-state on three qubit system and equivalent quantum walk scheme to obtain GHZ-state is illustrated. Red circle represents $\ket{0}$ of the real particle and green circle represents $\ket{1}$ of the real particle. Here the QW-based scheme is more simplified compared to quantum circuit implementation. The input state is $\ket{\Psi}_{in} = \ket{000}$ and output state is a GHZ-state $\ket{\Psi}_{out} = \frac{1}{\sqrt{2}}\Big( \ket{000} + \ket{111} \Big)$.}
	\label{cktmodelGHZ}
\end{figure}

\noindent
The scheme can help in reducing the time complexity for some circuits. One example is the circuit for preparing GHZ state, the complexity reduces by one step.  Fig.\,\ref{cktmodelGHZ} shows a three qubit circuit to create a GHZ-state and simplified implementation on quantum walk scheme. Notice that unlike the circuit model, which requires the application of three gates, the quantum walk can achieve the output in only two steps.\\
\noindent 
The quantum walk operation for creation of GHZ in computational basis is quite straightforward. For instance, a walker prepared in the state $\ket{000}$, upon being subjected to two steps of quantum walk can create a GHZ state. The first step would be the coin operator $\hat{C}(0,0,0,\pi/4 )$ followed by $S^1_+$ shift operator and the second step would be identity on coin state followed by  $S^1_+$ shift operator, where the coin operator $\hat{C}$ has been defined in equation\,\eqref{qcoin} and the $S^1_+$ in equation\,\eqref{qshift}. The sequence of steps, when executed will create the state $\frac{1}{\sqrt{2}}(\ket{000} + \ket{111})$.\\

\subsection{\label{sec:scaling} Scalability of the scheme}
\begin{figure}
	\centering
	\includegraphics[width=0.8\linewidth]{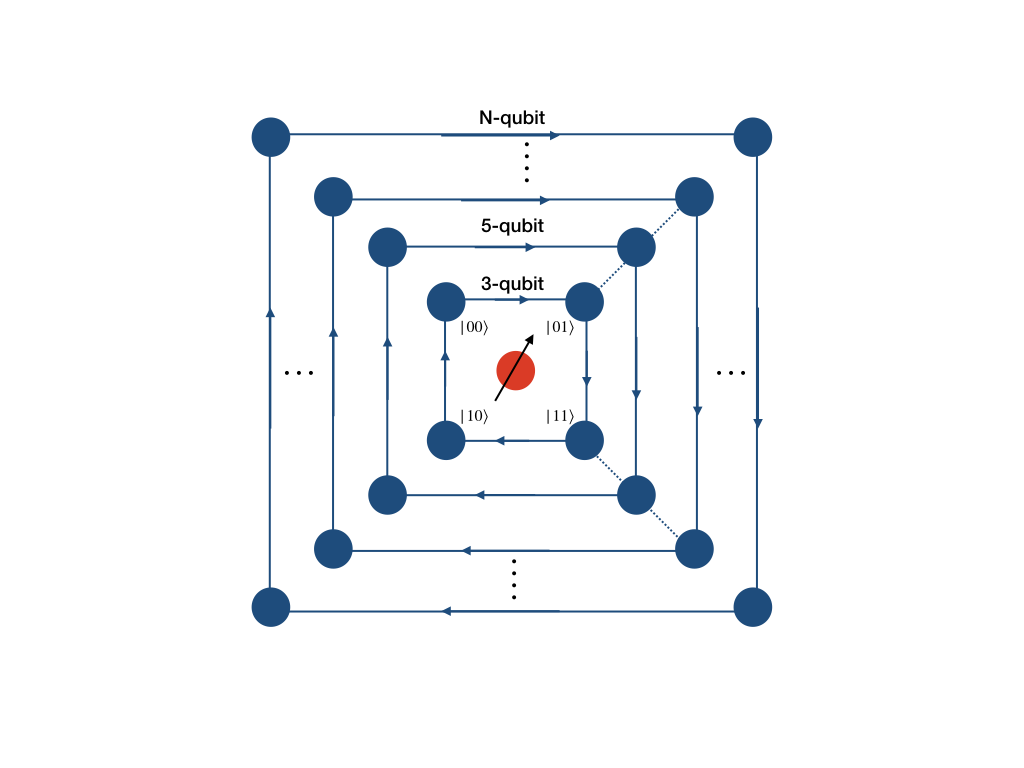}
	\caption{An illustration of scaling of three-qubit equivalent system to N-qubit system. Multiple closed graphs of four vertices equivalent to two qubit can be used to extended the quantum walk based universal quantum computer physically. Red solid circle is represents $\ket{0}$ of the real particle.}
	\label{MultiQ}
\end{figure}

\begin{figure}
	\centering
	\includegraphics[width=0.5\linewidth]{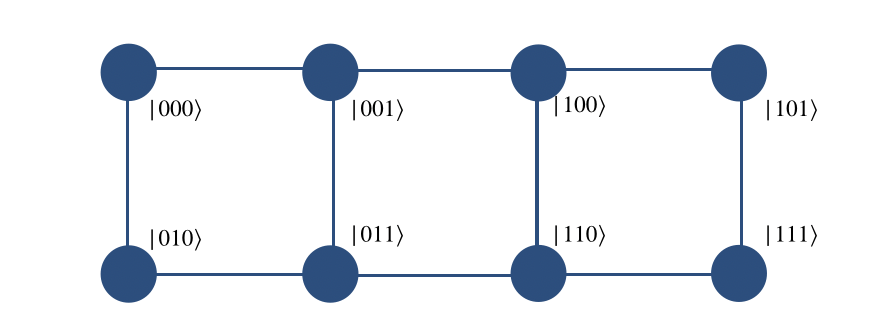}
	\caption{Illustration of an extension of position space mapping of three-qubit system to four-qubit system by connecting two three-qubit models position state by two edges. This is one of the many ways to multiply qubits for quantum walk based computational scheme.}
	\label{Map4qubit}
\end{figure}

\begin{figure}
	\centering
	\includegraphics[width=0.35\linewidth]{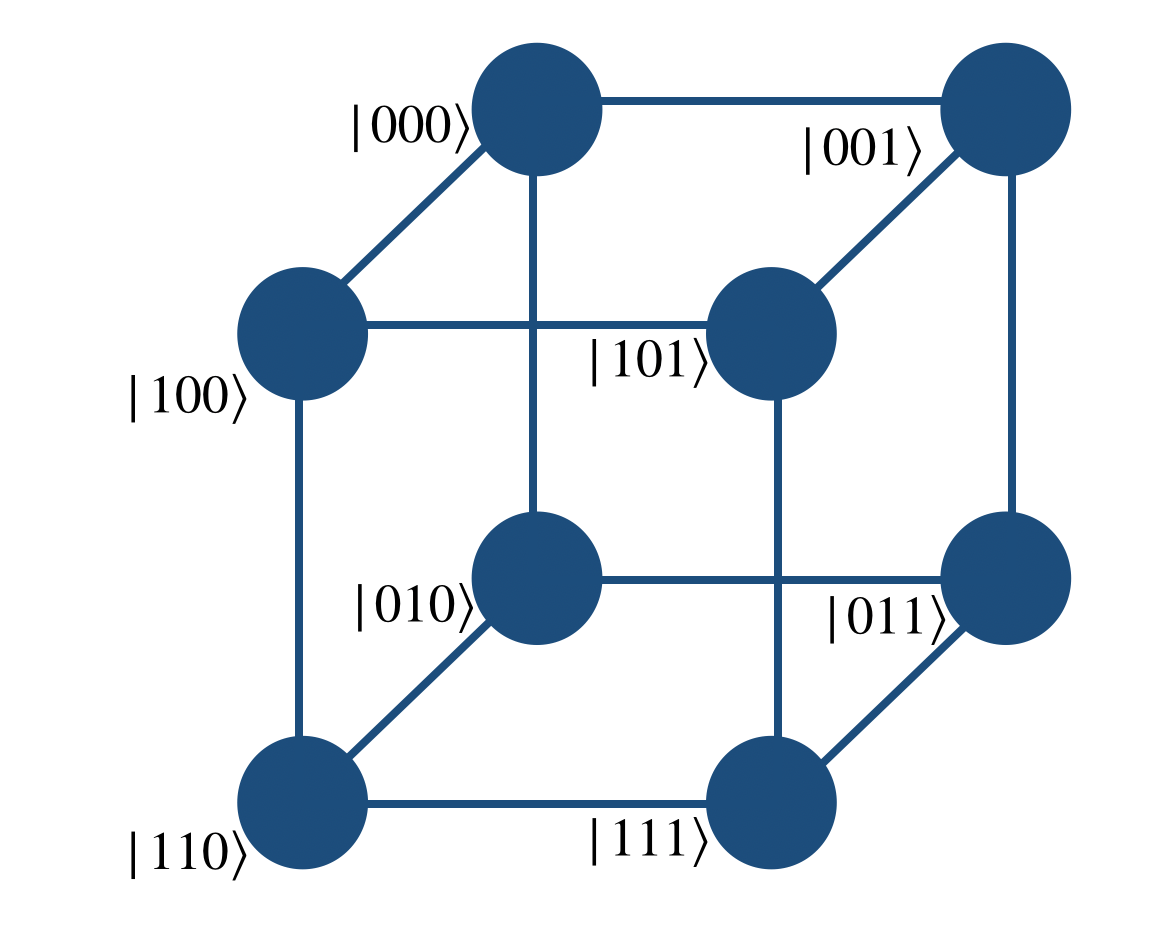}
	\caption{An illustration of the implementation of a three-qubit (eight point) position space for higher-qubit operations. }
	\label{4qubit}
\end{figure}

\noindent
The scheme can be scaled up to larger number of qubit by multiple ways.  Basic structure should follow Fig.\,\ref{Mapping} which systematically demonstrates the mapping of real qubit and its presence in superposition of position state to the multi-qubit computational basis. Extending the same scheme to higher dimensions to represent larger qubit systems is one way of extending the scheme to multi-qubit computation is one straight forward option. A single particle can perform universal computation on multi-qubit system with the help of multiple closed graphs of four vertices in tensor product as shown in Fig.\,\ref{MultiQ}. With an increase in number of qubits, different levels of two-qubit equivalent graphs can be added to the system. Each level communicates with different levels with the help of appropriate unitary evolution operators which are the extension of the operators presented for one-dimensional walk in Section\,\ref{sec:univcomp}. For example, if a gate is implemented on the fourth qubit, then walk is performed by the particle on the second level of the graph. In such case, we will apply identity on every other level and perform walk on the second level. E.g., if Hadamard operation is applied on the fourth qubit of the five qubit system, one will need two levels of closed graph such that the qubit state is given by $\ket{\phi_c} \otimes \ket{\phi_1} \otimes \ket{\phi_2}$. If the initial state of the walker is $\ket{00000}$ then the equivalent state on quantum walk scheme would be $\ket{0}_c \otimes \ket{00}_1 \otimes \ket{00}_2$, then applying $H_2$ from equation\,\eqref{HadaEq2} on the second level and identity on the first level will give Hadamard operation on fourth qubit,  
\begin{align}
H_4 \ket{0000} = (\mathbb{I}_1 \otimes W_{-}^{0})(H_1 \otimes \mathbb{I}_1 \otimes \mathbb{I}_2) .
\end{align}
Similarly, in this way, this scheme can be used to implement multi-qubit computation using quantum walk.  However, this scaling scheme is not unique and we can have different other possibilities of scaling on this scheme of computation. Some of the other possibilities are given in Figs.\,\ref{Map4qubit} and \ref{4qubit}. 

\noindent
Scaling onto a four-qubit system can be realized  by either considering two-particle quantum walk on a four position states, or a single-particle on an eight position state as shown in Fig.\,\ref{4qubit} and Fig.\,\ref{Map4qubit}, respectively. Similarly, a five-qubit system may be realized by a system with two-particle on an eight position state. In a system described as such, there will be two-particles and third qubit will be realized by superposition in the position space. An alternate realization of a five-qubit state can also be a three-particle quantum walk on a four position state.

\noindent
A subtle point to be made in using the eight position states with each state connected to three other state is that the set of shift operators required needs to be expanded to include operators that makes the particle choose a certain path. This is required because the set of universal gates can only act on a maximum of two qubit at a time and thus, each point in the position space must be connected to either one or two other states. Therefore, on a position space with eight states, the shift operator at each position must have three different variants which at a time can create the required superposition between two states in position space. Different architectures with well connected or limited connection can be further engineered to expand the scheme.  The configuration of the position space, connectivity and the ability to define the shift operators to transfer the particle across positions plays a crucial role in defining the operations of the larger qubits system.   

\subsection{\label{sec:Complexity} Quantum space complexity}
\noindent
The discrete-time quantum walk based scheme presented here only uses a single qubit in the form of quantum particle with two internal degrees of freedom on a graph. 
This scheme is physically equivalent to a three-qubit system reducing the space complexity by at least two-qubits on a three-qubit equivalent system. This scheme can be extended to $n$-qubit equivalent systems using the mapping between the position state and qubit state and hence reducing quantum space complexity. The extension scheme is not unique and it has many possibilities but a scheme with one-real qubit in the form of quantum particle on multiple layers (levels) of graph with four position states as shown in Fig.\,\ref{MultiQ} maps to $n-$qubit system and reduces the space complexity 
by up to $(n-1)$ qubits when compared to the standard circuit model implementation.  

\section{\label{sec:conc} Discussion and Conclusion}
\noindent
By using the provision of engineering the presence of a single particle in superposition of position space using discrete-time quantum walk, we have demonstrated the realization of universal quantum gates in a multi-qubit system. The main idea in this work is to demonstrate the effective use of controlled evolution of the particle on the position space and mapping the states of the system to the computational basis. We have  presented different constructions to show that the scheme can be scaled up to realize higher number of computational basis but an efficient scaling scheme needs some work. 
Scaling using combination of extended position space and a particle can be used to realize large dimensional computation basis.  For larger computational basis, if only one particle is considered, the position space required quickly scales up.  Therefore, a scheme of multi-particle quantum walk on extended position space could be more effective way to scale up the scheme.  Although the realization of Hadamard operation on computational basis looks a bit involved in the presented scheme, we can see that the gates like CNOT and Toffoli are more easily realizable.  However some realizable task shown here by this framework, are by no means exhaustive, and only provide a small glimpse into the possibilities of this scheme. With many experimental implementations of quantum walks in lattice based and photonic systems being reported, the idea from our scheme might motivate a new quantum computer architecture based on hopping of quantum particle in superposition of position space (lattice). Our scheme exploring the power of single particle in superposition of position space can immediately drive towards controlled engineering of quantum states and quantum simulations of sizable quantum system using fewer qubits. \\
\noindent
Quantum walks have been implemented on ion-traps\,\cite{BG02,HR09, FG10, RC12}, photonic systems\,\cite{MA10, AK10, AM10, PH14} and trapped atoms\,\cite{ML09, PR15}.    Quantum walk on a well-defined quantum systems with access to one-dimensional closed  graphs will be well suited to realize two and three qubit systems presented in this work. The experimental set-up for scaling needs multiple levels of one-dimensional closed graph to map to the many-particle states and position dependent evolution operators to implement quantum gates. Implementing any gate requires only few steps of quantum walk which are realizable. But implementing a complete circuit requires number of steps almost equivalent to the number of gates in the circuit. This implies that the scheme presented by us can be very well be used for quantum computation of small circuits implemented on near-term quantum devices.  Limitations with realizability of the required shift operators could be seen as an equivalent to the restricted connectivity we are seeing in the current available quantum processors. Increasing the size of the Hilbert space accessible without increasing the number of particle required for implementation is the key in the demonstrated protocol. 

\noindent
\section*{Acknowledgement}
\noindent
C.M.C. would like to thank Department of Science and Technology, Government of India for the Ramanujan Fellowship grant No.:SB/S2/RJN-192/2014 and the support from Interdisciplinary Cyber Physical Systems (ICPS) programme of the Department of Science and Technology, India, Grant No.:DST/ICPS/QuST/Theme-1/2019/1.

\noindent
\section*{Author contributions statement}
\noindent
C.M.C. designed the study, and S.S. and P.C. carried out the further work under the guidance of C.M.C. and prepared the figures. S.S., P.C., A.S. and C.M.C. together wrote the manuscript.


\end{document}